\documentclass[fleqn,12pt,twoside]{article}
\usepackage[headings]{espcrc1}
\readRCS
$Id: espcrc1.tex,v 1.2 2004/02/24 11:22:11 spepping Exp $ 
\usepackage{graphicx,psfig,epsfig}

\newcommand{\AmS}{{\protect\the\textfont2
  A\kern-.1667em\lower.5ex\hbox{M}\kern-.125emS}}

\def\la{\langle}
\def\ra{\rangle}
\def\beq{\begin{equation}}
\def\eeq{\end{equation}}
\def\be{\begin{eqnarray}}
\def\ee{\end{eqnarray}}

\def\k2av{\la k_T^2\ra}

\title{Di-hadron correlations at ISR and RHIC energies }

\author{
P. L\'evai\address[CNR]{Center for Nuclear Research, Department of Physics,\\
                       Kent State University, Kent OH-44242, USA}
\address[RMKI]{RMKI Research Institute for
        Particle and Nuclear Physics, \\
        P.O. Box 49, Budapest H-1525, Hungary},
G. Fai\addressmark[CNR]
\address[ELTE]{Department for Theoretical Physics,
        E{\"o}tv{\"o}s University, \\ 
        P{\'a}zm{\'a}ny P. 1/A, Budapest 1117, Hungary},  
and G. Papp\addressmark[ELTE]
\thanks{We thank Brian A. Cole for helpful discussions.
One of the authors (GF) acknowledges the support of a 
Szent-Gy\"orgyi Scholarship
of the Hungarian Department of Education.
This work was supported in part by  U.S. DOE grant DE-FG02-86ER40251,
NSF grant INT-0435701, and Hungarian grants T043455, T047050.}}

\begin{document}
\date{today} 
\maketitle

\begin{abstract}
The structure of hadron-hadron correlations is
investigated in proton-proton ($pp$) collisions.
We focus on the transmission of the
initial transverse momenta of partons
(``intrinsic $k_T$'') to the hadron-hadron correlations.
Values of the intrinsic transverse momentum obtained from
experimental correlations are compared to the results of a
model with partially randomized parton transverse momenta 
at ISR and RHIC energies. Procedures for extracting
the correlations from data are discussed.


\end{abstract}

\section{ Introduction}

Recent experimental data from the Relativistic Heavy Ion Collider (RHIC)
further emphasize the important role of jet physics in extracting 
information from relativistic proton-proton ($pp$) and nuclear
collisions about the strong interaction in the hard 
sector~\cite{PHENIXdijet,STARdijet,jqGLV,jqfai,jqBDMPS,jqQGP3}.
In particular, hadron-hadron correlations have been used to learn about the
transverse momentum distribution of partons in the 
proton~\cite{jjISR,E609,jjPHENIX,jia04,jjvitev}. 
Some of these measurements report a trigger dependent width for the 
intrinsic transverse momentum distribution of partons in the proton.
We aim to understand the origin of any such dependence, which appears
surprising at first sight.
The procedure to extract the transverse momentum width
of partons in the proton is complicated
experimentally because partonic properties need to be 
inferred from final-state hadrons.
On the theory side, perturbative quantum chromodynamics (pQCD)
studies have been carried out recently, primarily
with the goal of addressing asymmetries in polarized proton-proton 
scattering~\cite{Vogel04}. Here we restrict our attention to
unpolarized scattering and focus on the information
that can be obtained from measured quantities. 

In the framework of perturbative QCD, the starting point for the
description of
high transverse-momentum particle production in a $pp$ collision is 
provided by the secondary partons produced in 
elementary parton-parton collisions. 
Gluon radiation affects all partons, and the secondary partons 
may go through rescattering and multi-parton interaction
even in a $pp$ collision. Furthermore, the partons must finally hadronize, 
fragmenting into observable high-energy final-state hadrons. 
We address whether the width of the initial transverse
momentum distribution of the partons in the proton can be extracted
in this environment.


Another goal is to prepare the way for the study of 
hadron-hadron correlations
in nuclear collisions ($pA$, $dA$, and $AA$). Especially in matter,
while leading-order (LO) calculations have a clear physical
interpretation, there are many sources of complication in higher orders. 
In this paper we first consider $2 \rightarrow 2$ 
processes (LO) with an ``intrinsic'' transverse momentum. The 
average ``intrinsic'' $k_T$ effectively 
contains the contribution of perturbative soft gluon radiation, clouding
the separation of perturbative orders~\cite{Vogel04,Vitev05}.\marginpar{\it ?}
Further higher-order corrections include 
various random initial and final-state processes 
(elastic and inelastic scattering 
on colored scattering centers and hard radiation). 
Mirroring the experimental situation,
we focus on the transverse momentum components,
and find it economical for this purpose to represent these 
corrections by a randomization prescription. This way we maintain 
a simple interpretation of di-hadron correlations.
We show that the degree of randomization determines whether
the transverse-momentum width of the parton distribution in the proton 
can be extracted from data.
A naive procedure to accomplish this in the case of partial correlation
is outlined.

In the second part of the paper the complications from parton fragmentation
are introduced. We derive a practical formula to extract the
intrinsic transverse momentum width from the measured widths of the 
near-side and away-side peaks of the hadron-hadron correlation function.
The necessary approximations are discussed and the relation to 
commonly-used formulae is noted.

\section{Parton-parton correlation with intrinsic $k_T$ in $pp$ collision}
\label{sec_jjcorr}

The characteristics of the hadron-hadron correlations are inherited from
the parton level. 
\begin{figure}[htb]
\begin{center}
\resizebox{160mm}{70mm}{\includegraphics{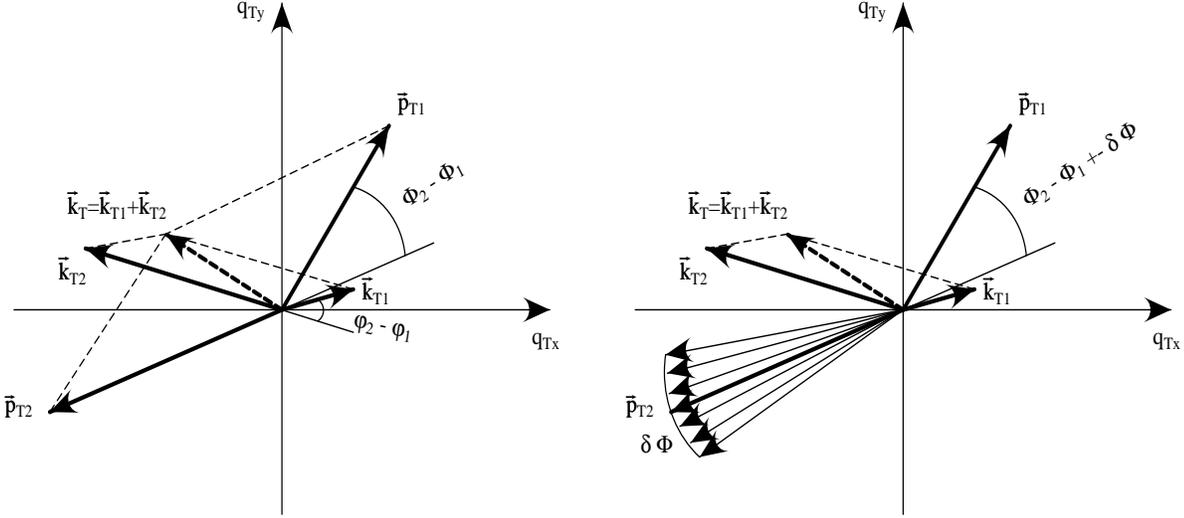}}
\end{center}
\vspace*{-0.5truecm}
\caption{Transverse kinematics of strongly-correlated parton-pair production
in a $2 \rightarrow 2$ reaction ({\sl left panel}).
Partially correlated  parton-pair production, where the
azimuthal angle, $\Phi_2$ of 
outgoing parton ``2'' shows a degree of 
randomization ({\sl right panel}). 
}
\label{fig1}
\end{figure}     
Figure 1 ({\sl left hand side}) displays the characteristics of a
simple $2 \rightarrow 2$ reaction in the transverse plane at the parton level.
The initial partons have transverse momentum components 
${\bf k}_{T1}$ and ${\bf k}_{T2}$, while the secondary partons are
characterized by transverse momenta ${\bf p}_{T1}$ and ${\bf p}_{T2}$,
respectively.
In general, transverse momentum conservation with initial and final
state radiation would take the form
\be
{\bf k}_{T1} + {\bf k}_{T2} + {\bf l}_{T1} + {\bf l}_{T2} + \ldots
= {\bf p}_{T1} + {\bf p}_{T2} +{\bf q}_{T1} + {\bf q}_{T2} + \ldots \ \ .
\label{imp00}
\ee
where ${\bf l}_{Ti}$ and ${\bf q}_{Ti}$ (i=1,2,...) 
represent transverse momentum transfers in initial and final state 
interactions. Since a complete treatments of 
this $n$-body problem is beyond our means, we start from
${\bf k}_{T1}, {\bf k}_{T2}$ and ${\bf p}_{T1}, {\bf p}_{T2}$ and attempt to 
include the corrections in a step-by-step process. In Sec. 2.1, we restrict
our attention to $2 \rightarrow 2$ reactions. In Sec. 2.2 we consider 
$2 \rightarrow 3$ ($3 \rightarrow 2$) reactions. The appearance of a $3^{rd}$ 
momentum leads to a partial loss of the strong correlations inherent in 
$2 \rightarrow 2$ reactions. In Sec. 2.3 
we introduce a model for the partially correlated case, where the loss of 
correlations can be a result of any number of further interactions.

\subsection{ $2 \rightarrow 2$ kinematics }

Let us consider an elementary $2 \rightarrow  2$ process. 
In this case transverse momentum conservation takes the form
\be
{\bf k}_{T1} + {\bf k}_{T2} = {\bf K}_{T} = {\bf p}_{T1} + {\bf p}_{T2} \ \ ,
\label{imp}
\ee
where $\bf K_T$ is the total transverse momentum of the parton pair.
This constraint results in a strongly correlated outgoing 
parton (and therefore jet) pair in the transverse plane.
Selecting outgoing parton ``1'' with transverse momentum of magnitude $p_{T1}$
and azimuthal angle $\Phi_1$ (which is random and uniformly distributed
in $[0,2\pi]$), the momentum
${\bf p}_{T2}$ is fully determined,  
i.e. both its magnitude $p_{T2}$ and azimuth $\Phi_2$ can be calculated.
In this fully determined system one can compute any correlation
between the outgoing partons ``1'' and ``2''. 
This situation will be referred to as ``strongly correlated'' ({\it ``sc''}).

For example, in the quantity 
$\langle (p_{T2} \sin(\Phi_2 - \Phi_1))^2 \rangle$
the averaging can not be split into
separate averaging of $p_{T2}^2$ and $(\sin(\Phi_2 - \Phi_1))^2$.
Instead:
\begin{eqnarray}
&& \langle (p_{T2} \sin(\Phi_2 - \Phi_1))^2 \rangle = 
{\cal A} + {\cal B} - {\cal C}  = \nonumber \\
&&\ \ =  \langle p^2_{T2} \sin^2 \Phi_2 \cos^2 \Phi_1 \rangle +
\langle p^2_{T2} \cos^2 \Phi_2 \sin^2 \Phi_1 \rangle
- 2 \langle p^2_{T2} \sin \Phi_2 \cos \Phi_1 \cos \Phi_2 \sin \Phi_1 \rangle
\ ,   \label{sindphi} \\
&& \langle (p_{T2} \cos(\Phi_2 - \Phi_1))^2 \rangle = 
{\cal D} + {\cal E} + {\cal C}  = \nonumber \\
&&\ \ =  \langle p^2_{T2} \cos^2 \Phi_2 \cos^2 \Phi_1 \rangle +
\langle p^2_{T2} \sin^2 \Phi_2 \sin^2 \Phi_1 \rangle
+ 2 \langle p^2_{T2} \cos \Phi_2 \cos \Phi_1 \sin \Phi_2 \sin \Phi_1 \rangle
\ ,   \label{cosdphi}
\end{eqnarray}
where we expect non-zero interference contributions (${\cal C}$) 
because of the strong correlation.

The direction of momenta ${\bf k}_{T1}$ and ${\bf k}_{T2}$
is described by azimuthal angles $\varphi_1$ and $\varphi_2$,
respectively, in the transverse plane of the collision,
in an arbitrary reference frame~\cite{Fn}.
We will assume that these angles are random and uniformly distributed
in $[0,2\pi]$, as is $\Phi_1$.  
The averaging over $\Phi_1, \varphi_1, \varphi_2$
can be carried out analytically and leads to:
\begin{eqnarray}
{\cal A}_{sc} = {\cal B}_{sc} & = & 
\frac{1}{8} p_{T1}^2 + \frac{1}{4} \left( k_{T1}^2 + k_{T2}^2 \right) 
\, \, \, ,
\hspace{2truecm} {\cal C}_{sc} =  \frac{1}{4} p_{T1}^2 
\, \, \, , \nonumber \\
{\cal D}_{sc} = {\cal E}_{sc} & = & 
\frac{3}{8} p_{T1}^2 + \frac{1}{4} \left( k_{T1}^2 + k_{T2}^2 \right) \ .
\label{ABCDE}
\end{eqnarray}

In the physical situation there is a distribution
of magnitudes for the initial transverse momenta,
${ k}_{T1}$ and ${ k}_{T2}$,
which appears in different theoretical
models of high-$p_T$ particle production in
proton-proton collisions~\cite{FieldFey,Wong98,Wang01,YZ02},
beyond the kinematics of eq.~(\ref{imp}).
Here we characterize
the transverse momenta of the initial partons (``intrinsic $k_T$'')
by a Gaussian distribution~\cite{Wang01,YZ02}:
\be
g(k_{Ti}) = \frac{1}{{2 \pi} \ \sigma^2_{k} } 
   \exp \left({-\frac{k_{Ti}^2}{2 \sigma^2_{k} }}\right) \ \ ,
\ \ \ \ \ i=1,2 \,\,\,  ,  \ \ 
\label{funcint}
\ee
where we assume that the width
$\sigma^2_{k}$ is the same for both distributions.
We expect this width,
including the effects of soft-gluon radiation, 
to be an intrinsic characteristic of proton
structure.  The two-dimensional width is defined as
$\langle k^2_T \rangle_{2D} = 2 \sigma^2_{k}$. 
In two-dimensional averaging 
$\langle k^2_T \rangle = \langle k^2_{Tx} \rangle + \langle k^2_{Ty} \rangle$,
and by symmetry
$\langle \vert k_{Ty} \vert \rangle = \sqrt{\langle k^2_T \rangle/ \pi}$.

Numerically computing the correlations in eqs.~(\ref{sindphi})-(\ref{cosdphi})
with the distribution (\ref{funcint}) yields
\begin{eqnarray}
{\cal A}_{sc} = {\cal B}_{sc} &=& 
\frac{1}{8} p_{T1}^2  + \sigma^2_{k} \,\,\,   ,
\hspace{2truecm}
{\cal C}_{sc} = \frac{1}{4} p_{T1}^2  \,\,\,   , \nonumber \\
{\cal D}_{sc} =
{\cal E}_{sc} &=& \frac{3}{8} p_{T1}^2 + \sigma^2_k  \,\,\,   .
\end{eqnarray}
Summing these contributions gives
\begin{eqnarray}
\langle (p_{T2} \sin(\Phi_2 - \Phi_1))^2 \rangle_{sc} &=& 
   2 \cdot \sigma^2_k  \ \ ,            \label{sinsc} \\
\langle (p_{T2} \cos(\Phi_2 - \Phi_1))^2 \rangle_{sc} &=& 
   p_{T1}^2 + 2 \cdot \sigma^2_k \ \ . \label{cossc}
\end{eqnarray}

Thus, in a strongly correlated parton system
correlation (\ref{sinsc}) directly displays the width of the
intrinsic transverse momentum distribution without any dependence
on other variables. This correlation is sought experimentally.

\subsection{ $2 \rightarrow 3$  ($3 \rightarrow 2$) kinematics}

While eq.~(\ref{imp}) is completely general for a $2 \rightarrow  2$ 
partonic process, the presence of a  $3^{rd}$ non-zero momentum 
(initial or final) 
will destroy transverse momentum conservation for the two outgoing 
partons and weaken their correlation.
In this case one has  ${\bf K'}_{T} = {\bf K}_{T} + {\bf l}_{T1}$
or ${\bf K'}_{T} = {\bf K}_{T} - {\bf q}_{T1}$. As long as no variation is 
allowed for the magnitudes $k_{Ti}$ (i=1,2), a calculation analogous to 
(\ref{imp})-(\ref{cosdphi}) can be carried out analytically with a result
similar to (\ref{ABCDE}).
However, when a distribution for $k_{Ti}$ is considered, one needs to resort to 
a numerical evaluation leading to  
 \begin{eqnarray}
\langle (p_{T2} \sin(\Phi_2 - \Phi_1))^2 \rangle_{sc} &=& 
   2 \cdot \sigma^2_k + \Sigma^2  \ \ ,            \label{sin3} \\
\langle (p_{T2} \cos(\Phi_2 - \Phi_1))^2 \rangle_{sc} &=& 
   p_{T1}^2 + 2 \cdot \sigma^2_k + \Sigma^2 \ \ , \label{cos3}
\end{eqnarray}
with an additional term $\Sigma^2$ compared to (\ref{cossc}). 
If the magnitude of the extra momentum ${\bf p}_{T3}$ (= $q_{T1}$ or $l_{T1}$)
is uniformly distributed in $[0,p_{Tmax}]$, the additional term is 
$\Sigma^2 = p_{Tmax}^2/6$. If, on the other hand, the extra momentum is 
limited in magnitude by $K_T$ (due to e.g. transverse energy conservation 
in the model), then $\Sigma^2 = 2 (2\sigma_k^2) p_{Tmax}^2/6$, reflecting the
variation of $k_{Ti}$.

It is cumbersome to consider the physical nature of the additional momenta 
${\bf l}_{T1}, {\bf l}_{T2}, \ldots,$ ${\bf q}_{T1}, {\bf q}_{T2}, \ldots$
explicitely. Fortunately, since we want to maintain an interpretation
centered around ${\bf k}_{T1}, {\bf k}_{T2}$ and ${\bf p}_{T1}, {\bf p}_{T2}$,
this is not required. It is sufficient to focus attention on the  
$2 \rightarrow 2$ process, taking into account the effect of all other 
complications by allowing a randomness in the momenta of the final-state 
partons. We next turn to a model based on this picture.
 
\subsection{Partial correlation by randomization}

The effect of the third momentum will be described on the average
as a randomizing 
change on the final momenta of the $2 \rightarrow  2$ process.
The randomness introduced by this and other higher-order effects
could lead to fully independent
and uniformly distributed azimuthal angles $\Phi_1$ and $\Phi_2$
for the outgoing partons.
In this case the interference term becomes zero, ${\cal C}_{rand}=0$, 
and all quadratic terms are equal, 
${\cal A}_{rand} = {\cal B}_{rand} = 
 {\cal D}_{rand} = {\cal E}_{rand} = \langle p^2_{T2}\rangle /4$. 
These results satisfy the expectation for a fully randomized system:
\be
\langle (p_{T2} \sin(\Phi_2 - \Phi_1))^2 \rangle_{rand}
= \langle (p_{T2} \cos(\Phi_2 - \Phi_1))^2 \rangle_{rand} = 
\langle p^2_{T2}\rangle /2  \,  . \label{sinrand}
\ee
The  expectation value (\ref{sinrand}) of parton ``2'' is related 
by symmetry to $\langle p^2_{T1} \rangle$ as
$\langle p^2_{T2} \rangle \equiv \langle p^2_{T1} \rangle $.
In this limit
all information about initial $k_T$ and 
$\sigma_k$ has been lost and can not be recovered from
final hadron-hadron correlations.

However, an entire spectrum of intermediate scenarios exists 
between the strongly correlated and the fully randomized cases
with partial loss of the strong correlation.
Let us assume that outgoing parton ``2'' suffers  partial
randomization. Thus, for example, the azimuthal angle $\Phi_2$ 
acquires a distribution around its strongly correlated
value, determined 
by the $2 \rightarrow 2$ collision, see Figure 1 ({\sl right 
hand side}). This randomization takes place in a cone around the 
original well-determined direction.

We examine two types of randomization in the transverse plane:
\begin{description}
\item{A)} Gaussian randomization with a one-dimensional weight function: 
\be
w(\Phi^*_2) = \frac{1}{\sqrt{2 \pi} \ \sigma_\Phi} 
         \exp{\left( -\frac{(\Phi^*_2-\Phi_2)^2}{2 \sigma_\Phi^2}\right)} \ \ ;
\ee
\item{B)} uniform randomization in the azimuth region
$[\Phi_2 - \delta \Phi, \Phi_2 + \delta \Phi]$.
\end{description}

Our aim in this Section is to determine the influence of these 
randomization assumptions on the basic $2 \rightarrow 2$ process.
We re-calculate the quadratic and interference 
terms in eqs.~(\ref{sindphi})-(\ref{cosdphi}).  
For simplicity, we introduce the following notation:
\begin{eqnarray}
\alpha &=& 4 \, \left[ ({\cal A} + {\cal B} - {\cal C} ) 
          - 2 \cdot \sigma^2_k \right] / p_{T1}^2 \,\,\, , \\
\beta  &=& 4 \, \left[ ({\cal D} + {\cal E} + {\cal C} )
          - 2 \cdot \sigma^2_k \right] / p_{T1}^2 \,\,\, , \\
\gamma &=& 4 \, {\cal C}  / p_{T1}^2 \,\,\, .
\end{eqnarray}
According to eqs.~(\ref{sindphi}) and (\ref{cosdphi}),
coefficients $\alpha$ and $\beta$  give a clear account of
the $p_{T1}$ dependence of
the correlations in the partially correlated (``{\sl pc}'') case:
\begin{eqnarray}
\langle (p_{T2} \sin(\Phi_2 - \Phi_1))^2 \rangle_{pc} &=&
\alpha \left( \frac{p_{T1}}{2}\right)^2 + 2 \cdot \sigma^2_k
\label{partonpc1} \,\,\, , \\
\langle (p_{T2} \cos(\Phi_2 - \Phi_1))^2 \rangle_{pc} &=&
\beta \left( \frac{p_{T1}}{2}\right)^2 + 2 \cdot \sigma^2_k  \,\,\, .
\label{partonpc}
\end{eqnarray}
This $p_{T1}$ dependence was completely absent in eq.(\ref{sinsc}) for
$\langle (p_{T2} \sin(\Phi_2 - \Phi_1))^2 \rangle_{sc}$.
The value of coefficient $\gamma$ indicates the level of correlation.

Figure 2 displays our numerical results as functions of
$\sigma_{\Phi}$ for case A ({\sl solid lines}) and $\delta \Phi$ for case B
({\sl dashed lines}).  We found $\alpha$, $\beta$ and $\gamma$
to be independent of $\sigma_k$ and $p_{T1}$.
They depend only on $\sigma_\Phi$ in case A, and on $\delta \Phi$ in case B.
The partially correlated system turns into a fully randomized one
when the interference term disappears and the quadratic sine and cosine
terms become equal.
This happens at the same point in both cases, namely
at $\sigma_\Phi = \pi/2$ and at $\delta \Phi = \pi/2$.
The result for case B confirms the expectation that complete 
loss of correlation
occurs when the opening angle of the randomization cone becomes $\pi$. 
The shape of the curves slightly depends on the nature of 
the microscopic processes represented in our picture by the 
different randomization prescriptions.
\begin{figure}[htb]
\begin{center}
\resizebox{92mm}{92mm}{\includegraphics{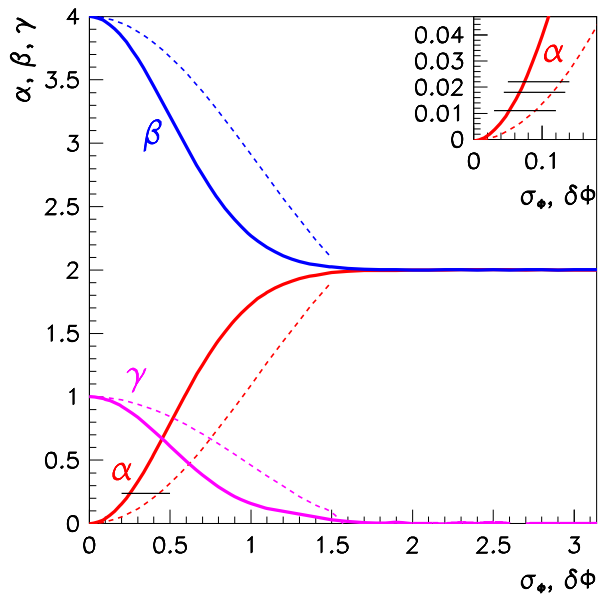}}
\end{center}
\vspace*{-1.0truecm}
\caption{
The coefficients $\alpha$, $\beta$ and $\gamma$ as functions of
$\sigma_\Phi$ in case A ({\sl solid lines}) 
and as functions of $\delta \Phi$ in case B ({\sl dashed lines}).
A horizontal line in the lower left corner indicates the value of $\alpha$ 
in $pp$ collisions at RHIC energy, $\sqrt{s}=200$ GeV.
In the upper right a magnified view of the small $\alpha$ region is given.
The horizontal lines indicate the values of $\alpha$ at ISR
energies (see Section 3 and Table 1).
}
\label{fig2}
\end{figure}     

The effect of these fluctuations (randomization) at the parton collision
level will be superimposed on the intrinsic transverse momentum of 
partons expressed by $\sigma_k$. 
Furthermore, 
hadronization (jet fragmentation) will complicate the 
picture. We incorporate the consequences of hadronization gradually: 
in a first step, in Section~\ref{sec_pc}, the momentum fraction of the
trigger hadron is taken into account.
Section~\ref{sec_hh} contains a more detailed discussion.

\section{Partial correlations at ISR and RHIC energies}
\label{sec_pc}

Data on di-hadron correlations in $pp$ collisions at
ISR energies $\sqrt{s} = 31, \ 45, \ 62$~GeV~\cite{jjISR}
and at RHIC energies $\sqrt{s} = 200$ GeV~\cite{PHENIXdijet,jjPHENIX}
show a dependence on the transverse momentum of the trigger
hadron $p_{T,trig}$, which is at this point in our treatment
represented by $p_{T1}$. 
Figure 3 displays available $pp$ data at ISR and RHIC.
\begin{figure}[htb]
\begin{center}
\resizebox{75mm}{80mm}{\includegraphics{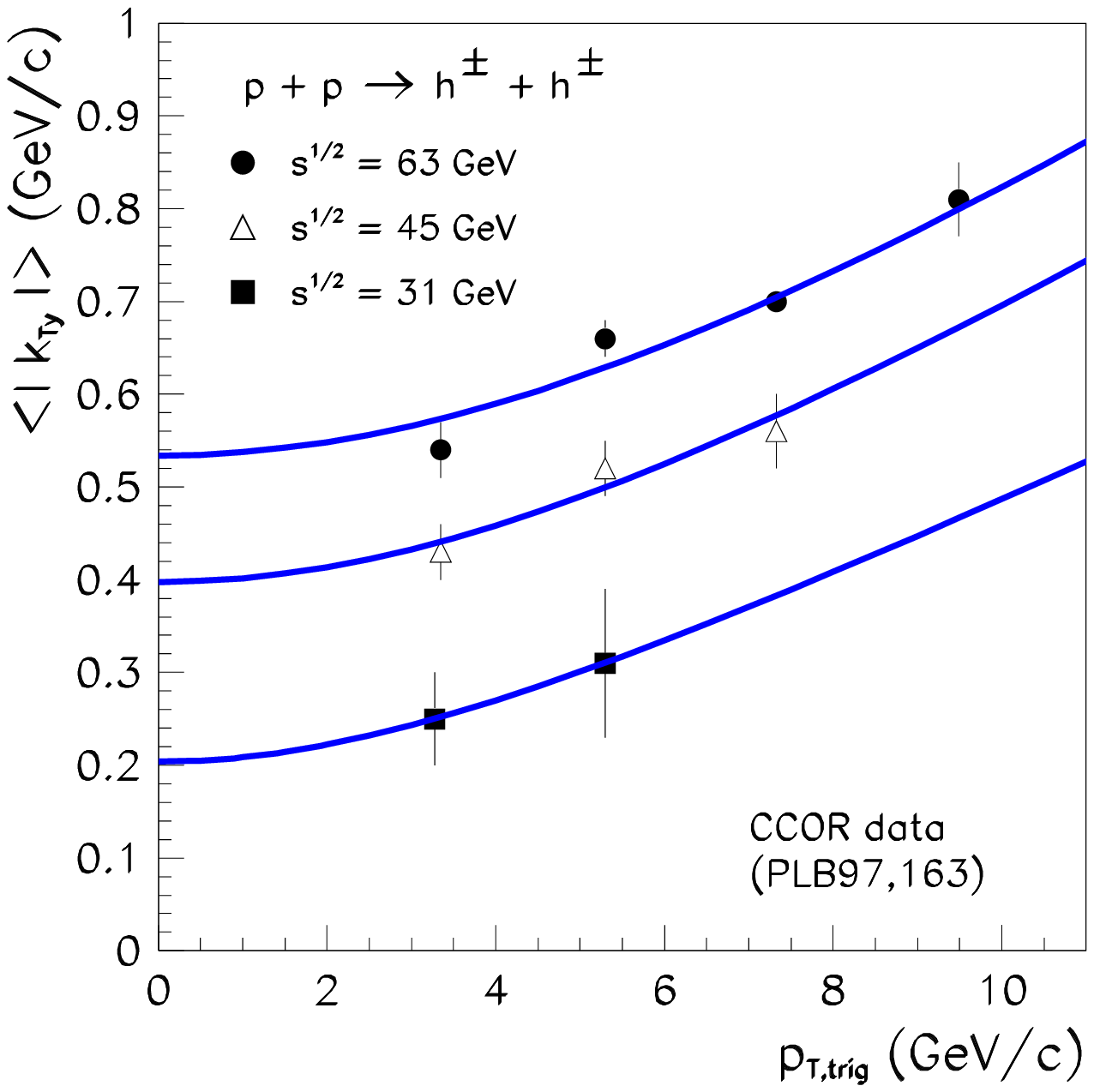}}
\resizebox{82mm}{80mm}{\includegraphics{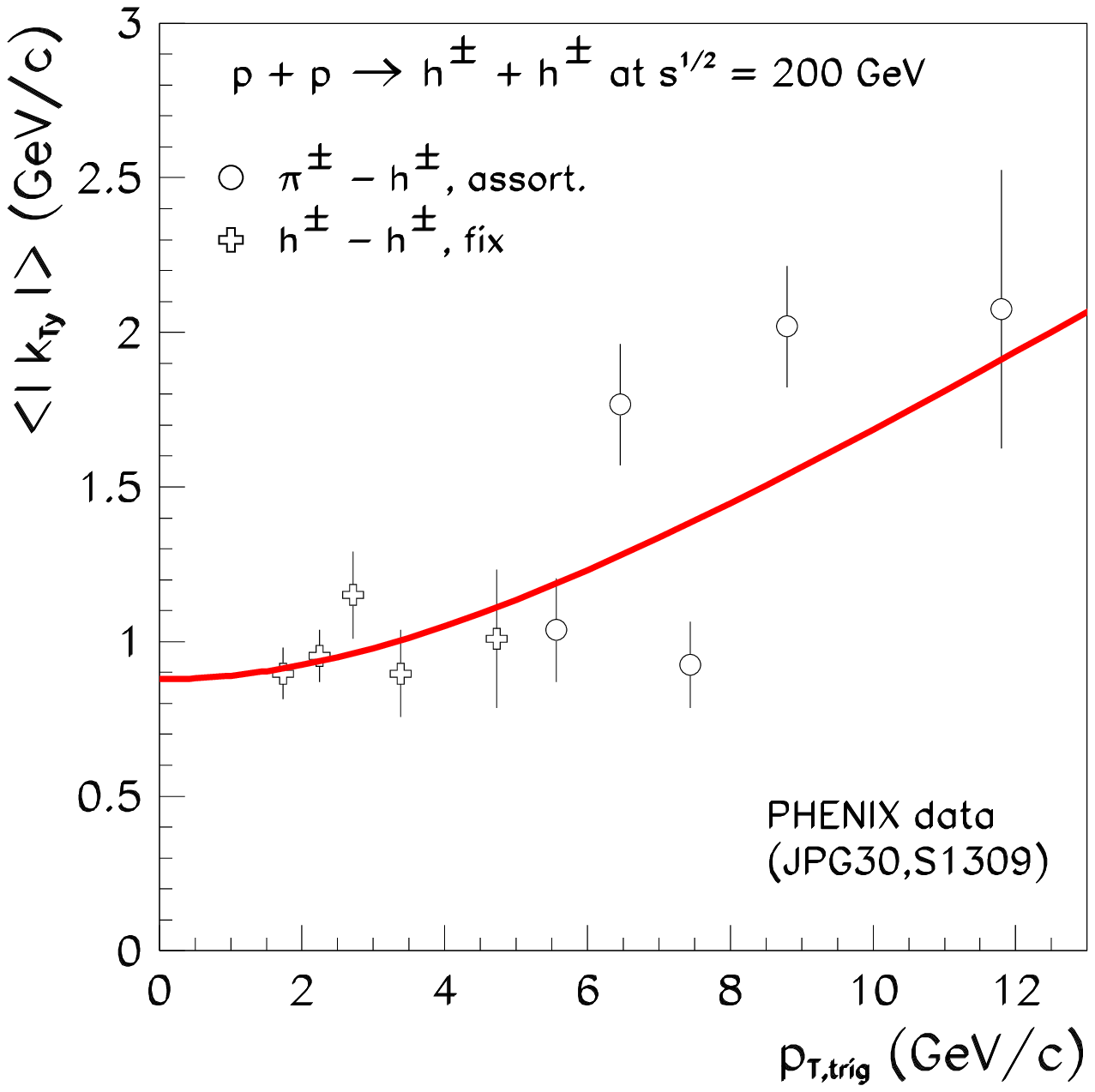}}
\end{center}
\vspace*{-0.7truecm}
\caption{
Data published for hadron-hadron correlations in
$pp$ collisions at ISR~\cite{jjISR} ({\sl left panel})
and RHIC~\cite{jjPHENIX} ({\sl right panel}) together with the fit for
a partially correlated parton system (see text and Table 1.).
}
\label{fig3}
\end{figure}     

Since the  parton-level quantity
$\langle \vert k_{Ty} \vert \rangle  =
\sqrt{(\langle (p_{T2} \sin(\Phi_2 - \Phi_1))^2 \rangle)/\pi }$
is independent of $p_{T1}$ in a strongly correlated system
(see eq.(\ref{sinsc})),
the $p_{T1}$ dependence on Fig. \ref{fig3} indicates a
partially correlated situation at ISR and RHIC (see eq.(\ref{partonpc1})):
\be
\langle \vert k_{Ty} \vert \rangle  =
\frac{1}{\sqrt{\pi}}
\sqrt{\langle (p_{T2} \sin(\Phi_2 - \Phi_1))^2 \rangle_{pc} }
= \frac{1}{\sqrt{\pi}}
\sqrt{\frac{\alpha}{4} p^2_{T1} + 2 \cdot \sigma^2_k}  \,\,  ,
\label{corrcoef}
\ee
meaning that the intrinsic width $\sigma_k$ can be identified as the
limiting value when $p_{T,trig} \rightarrow 0$.

This expression is valid if parton fragmentation is neglected.
To include fragmentation effects at this stage
we use a simple prescription to take into account that 
the experimentally measured trigger hadron carries a 
fraction of the corresponding parton transverse momentum,
and we correct with the mean 
momentum fraction carried by the produced hadron,
\be
p_{T,trig} = z \cdot p_{T1} \,\,\, . 
\label{ztrig}
\ee
Thus, in terms of hadronic variables the expression (\ref{corrcoef}) 
takes the following form:
\be
\langle \vert k_{Ty} \vert \rangle
= \frac{1}{\sqrt{\pi}}
\sqrt{\frac{\alpha}{4 z^2} p^2_{T,trig} + 2 \cdot \sigma^2_k}  
\,\,  .
\label{corrcoefh}
\ee 

Eq.~(\ref{corrcoefh}) can be compared to existing experimental data.
For RHIC energy the fragmentation correction is
$\langle z \rangle_{p_T > {\textrm{3\ GeV}} }
= 0.75 \pm 0.05$~\cite{jjPHENIX,faix}.
For ISR data we use $\langle z \rangle \approx 0.85$~\cite{jjISR,jjPHENIX}.
Since $\langle z \rangle$ is independent of $p_{T,trig}$~\cite{jjPHENIX},
it can be easily included in eq. (\ref{corrcoefh}) by rescaling $\alpha$.
Figure 3 shows the best fit according to eq. (\ref{corrcoefh}).
Table 1 summarizes the parameters of the fit.

\begin{table}[htb]

\caption{Extracted correlation coefficient, $\alpha$, and the
width of the initial parton transverse momentum distribution, $2 \sigma^2_k$,
at ISR and RHIC energies in $pp$ collisions.}
\label{table:1}

\newcommand{\m}{\hphantom{$-$}}
\newcommand{\cc}[1]{\multicolumn{1}{c}{#1}}
\renewcommand{\tabcolsep}{2pc} 
\renewcommand{\arraystretch}{1.2} 
\smallskip
\begin{tabular}{@{}cccc}

\hline

$\sqrt{s}_{pp}$  (GeV) &$\langle z \rangle $

     & $\alpha$ & $2~\sigma^2_k$ (GeV$^2$) \\ \hline

31 &  0.85  & 0.011  &0.13  \\  \hline

45 &  0.85  & 0.018  &0.50  \\  \hline

62 &  0.85  & 0.022  &0.90  \\  \hline

200 & 0.75  & 0.240  &2.42   \\  \hline

\hline
\end{tabular}\\[2pt]
\vspace{-5mm}
\end{table}

It is of interest to remark here that
the obtained width at RHIC energy matches closely the value 
used in a transverse-momentum augmented
perturbative QCD calculation in $pp$
collisions in Ref.~\cite{dau_levai03}:
$ 2 \, \sigma_k^2 = 2.5$ GeV$^2$.
As discussed earlier, this value contains the effect of
initial soft gluon radiation.
In nuclear collisions ($dAu$ and $AuAu$), the widths $\sigma_k$ can be
different, in which case $2 \, \sigma_k^2 
\rightarrow \sigma_{k1}^2 + \sigma_{k2}^2$.

However, parton fragmentation has a more complicated effect
on the relation between measurable quantities and parton 
transverse momenta than reflected in eq. (\ref{ztrig}). Next we
examine to what extent the results in Table~1 and Fig.~3 remain
valid in a more realistic treatment of hadronization.

\newpage

\section{Hadron-hadron correlations in $pp$ collisions}
\label{sec_hh}

In a real $pp$ experiment, of course, hadron-hadron correlations are measured. 
This introduces several complications in the evaluation of 
correlations
and of parton intrinsic transverse momenta. 
Fragmentation of the secondary partons produces a distribution of
outgoing hadrons. The experiments pick a trigger hadron with
transverse momentum $p_{T,trig}$ and an associated hadron with
transverse momentum $p_{T,assoc}$ from the measured 
distribution. The azimuthal correlation function displays a two-peak 
structure, where the width of the near-side peak is denoted by 
$\sigma_N$ and the width of the away-side peak is $\sigma_A$.
The value of $\sigma_N$ carries information on the fragmentation 
process only, while $\sigma_A$ may contain the contribution of the
intrinsic transverse momentum. In a fully randomized situation 
all information about intrinsic $k_T$ is lost (see eq.(\ref{sinrand}))
and we would expect $\sigma_A \approx \sigma_N$. However, measurements
show $\sigma_A > \sigma_N$, indicating that the width of the intrinsic
transverse momentum distribution can be extracted from hadron-hadron
correlations. 
\begin{figure}[htb]
\begin{center}
\resizebox{160mm}{80mm}{\includegraphics{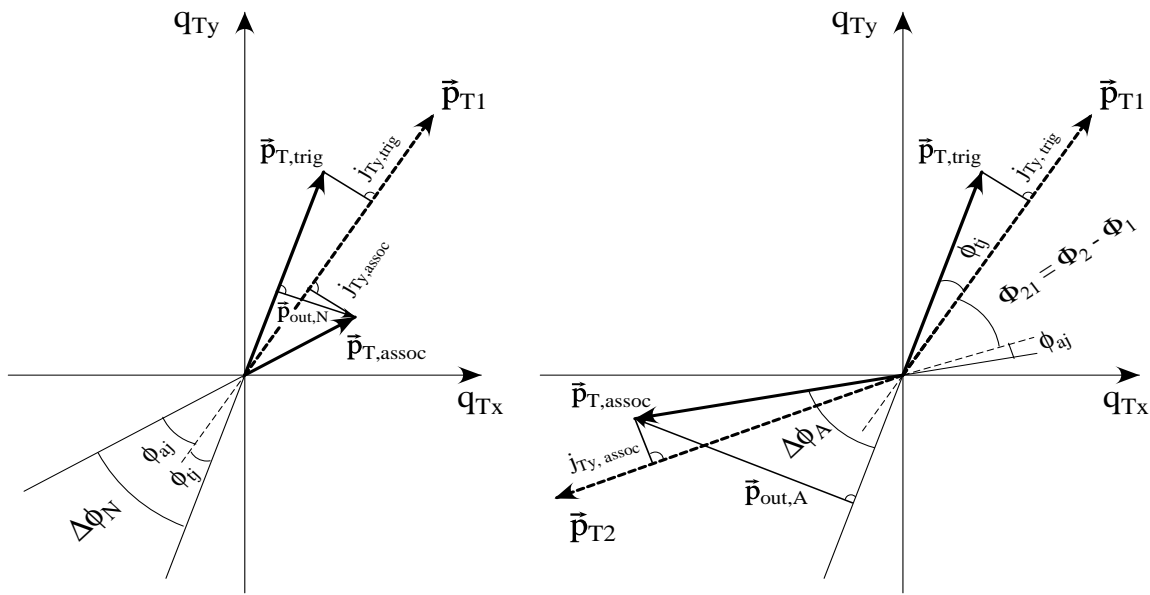}}
\end{center}
\vspace*{-1.0truecm}
\caption{
Schematic view of the near side ({\sl left panel}) and away side 
({\sl right panel})
correlations with parton fragmentation and notation for angles 
(see text).
}
\label{nearaway}
\end{figure}

\subsection{Near-side correlation and parton fragmentation}
\label{sec_near}

Data from RHIC show that in $pp$ and $dAu$ collisions 
the widths of the near-side peaks, $\sigma_{N}$,
are in good agreement~\cite{jjPHENIX,jia04}.
An average value for the transverse momentum component 
of the trigger hadron relative to the jet axis,
$\langle \vert j_{Ty} \vert \rangle = 324 \pm 6$~MeV/c has been
obtained in the usual notation~\cite{Fn}. 
This value agrees within error bar with the $AuAu$ 
value at RHIC, independent of centrality~\cite{jjPHENIX},
which indicates a collision-system independent fragmentation. 
Furthermore, the data are close to the ISR value for $pp$ collisions, 
$\langle \vert j_{Ty} \vert \rangle = 393 \pm 3$ MeV/c~\cite{jjISR}. 
These results suggest the existence of an approximately
universal fragmentation pattern,
almost independent of energy, centrality and collision system.

Here we summarize the assumptions needed
to obtain an expression for the near-side
azimuthal correlation. 
Figure \ref{nearaway} ({\sl left side})
illustrates parton fragmentation and the definition of the 
quantities necessary to extract the near-side correlation.

In near-side correlations, the
trigger hadron and the associated hadron both arise from the same parton and
the following relation holds: 
$\sin \Delta \phi_N = \sin (\phi_{tj} + \phi_{aj})$.
In light of our analysis in previous Sections, we calculate
$\langle p_{out,N}^2 \rangle = 
\langle (p_{T,assoc} \sin \Delta \phi_N)^2 \rangle$.
Since fragmentation angles $\phi_{tj}$ and $\phi_{aj}$ are 
statistically independent, we obtain the following expression
for averaged azimuths, where the cross terms  have been dropped
(since they average to zero):
\be
\langle {p^2_{out,N}} \rangle = 
\langle p_{T,assoc}^2 \sin^2 \Delta \phi_N \rangle =
\langle p_{T,assoc}^2 \sin^2 \phi_{tj} \cdot \cos^2 \phi_{aj} \rangle +
\langle p_{T,assoc}^2 \cos^2 \phi_{tj} \cdot \sin^2 \phi_{aj} \rangle .
\label{nearsin}
\ee
Following the notation of Fig. 4, 
introducing $x_h = p_{T,assoc}/p_{T,trig}\,$, and substituting the azimuths
in eq.~(\ref{nearsin}):
\begin{eqnarray}
\langle p^2_{out,N}  \rangle &=&
\left\langle j^2_{Ty,trig} \, x_h^2 \,    
\left( 1 - \frac{j^2_{Ty,assoc}}{p^2_{T,assoc}}  \right) \right\rangle
+ \left\langle 
\left( 1 - \frac{j^2_{Ty,trig}}{p^2_{T,trig}}  \right) 
j^2_{Ty,assoc}  \right\rangle \, .
\label{neareq1}
\end{eqnarray}

For $p_{T,trig}, \ p_{T,assoc} \geq 2$ GeV we can neglect the
terms $j^2_{Ty,assoc}/p^2_{T,assoc} \sim {\cal O}(0.05)$
and $j^2_{Ty,trig}/p^2_{T,trig} \sim {\cal O}(0.05)$,
generating a 10 \% uncertainty in the results. 
Eq.~(\ref{neareq1}) simplifies to
\be
\left\langle p^2_{out,N} \right\rangle =
\langle j^2_{Ty,trig} \,  x^2_h \rangle
+ \langle j^2_{Ty,assoc} \rangle \ . \label{neareq2}
\ee
Assuming a universal fragmentation pattern
characterized by $\langle j^2_{Ty} \rangle \equiv
\langle j^2_{Ty,assoc} \rangle \approx \langle j^2_{Ty,trig}\rangle$ and
$x_h$ independence, one gets
\be
\left\langle p^2_{out,N} \right\rangle = 
\langle j^2_{Ty} \rangle
\cdot (1+ \langle x^2_h \rangle) \ . \label{neareq3}
\ee
In a final step we connect 
$\left\langle p^2_{out,N} \right\rangle$ to the measured width 
of the near side azimuthal correlation.
Assuming a Gaussian distribution and following the usual procedure~\cite{jia04},
one can approximate $\left\langle p^2_{out,N} \right\rangle$
within 10 \% precision with the following expression:
\begin{eqnarray}
\langle p^2_{out,N} \rangle &\approx&
\left\{ \begin{array}{lll} \langle p^2_{T,assoc} \rangle \ \sigma^2_N &&
\mbox{\rm if \hspace*{.4 cm} $\Delta \phi_N \in [0,\pi/8 ]$}  
\\
\langle p^2_{T,assoc} \rangle [ \sin \sigma_N^2 - \sigma_N^4]  &&
\mbox {\rm if \hspace*{.4 cm} $\Delta \phi_N \in 
\left[\pi/8,\pi/5 \right]$}   
\end{array}  \right . \,\,\, .
\label{neareq4}
\end{eqnarray}
Experimental data for the near side correlation
at ISR and RHIC energies~\cite{jjISR,jjPHENIX} allow us to apply
the approximation in the top line of eq.~(\ref{neareq4}). 
Since in two-dimensional averaging for any quantity $X$ we have
$\langle \vert X \vert \rangle^2 =
2/\pi \cdot \langle X^2 \rangle$, one can simplify
eq.(\ref{neareq3}) and obtain the usual form~\cite{jjISR,jjPHENIX,jia04} 
for the fragmentation broadening:
\be
\langle \vert j_{Ty} \vert \rangle =
\frac{\sqrt{2 \langle p^2_{T,assoc} \rangle}}{\sqrt{\pi}} \ 
\frac{\sigma_N}{\sqrt{1+\langle x_h^2 \rangle}}
\label{nearfin}
\ee
This can be evaluated from  
data on $p_{T,assoc}$, $p_{T,trig}$ and 
the near-side azimuthal width $\sigma_N$.
The result for $\langle \vert j_{Ty} \vert \rangle$
does not depend on $p_{T,trig}$ directly, the only
dependence comes through the quantity $x_h$. 
We are now in a position to investigate the
away-side correlation and its relation to the
intrinsic transverse momentum.

\subsection{Away-side correlation and intrinsic transverse momentum}
\label{sec_away}

Away-side hadron correlations carry information on 
the intrinsic transverse momentum of partons.
Fig. \ref{nearaway} ({\sl right side}) 
displays the geometry of this correlation.
Here, the trigger hadron and the associated hadron originate in
different partons characterized by different transverse momenta
${\bf p}_{T1}$ and ${\bf p}_{T2}$, respectively.
We choose ${\bf p}_{T1}$ to be the transverse momentum of the parton which
will give birth to the trigger hadron with momentum ${\bf p}_{T,trig}$.

We need to evaluate the expectation value of
$p^2_{out,A} =  p^2_{T,assoc} \sin^2 \Delta \phi_A $ 
to determine the away-side azimuthal correlation. One can
read off of Fig.~4 ({\sl right side})
that $\Delta \phi_A = \phi_{tj} + \Phi_{21} + \phi_{aj}$,
where $\Phi_{21} = \Phi_2 - \Phi_1$.
All cross terms drop in the averaging, since fragmentation
and initial kinematics, and fragmentation of trigger and associated
hadrons are independent.

Now one can express the wanted $\langle p^2_{out,A} \rangle$ as
\begin{eqnarray}
\langle p^2_{out,A} \rangle  
&=& \langle p^2_{T,assoc} \sin^2 (\phi_{tj} + \Phi_{21} + \phi_{aj}) \rangle =
\nonumber \\
&=& \langle p^2_{T,assoc} (1- \sin^2 \Phi_{21}) \ 
(\sin^2 \phi_{tj} \cos^2 \phi_{aj} + \cos^2  \phi_{tj} \sin^2 \phi_{aj}) 
\rangle + \nonumber \\  
&& + \langle p^2_{T,assoc} \sin^2 \Phi_{21} \ 
(1 - \sin^2 \phi_{tj} \cos^2 \phi_{aj} - 
     \cos^2  \phi_{tj} \sin^2 \phi_{aj}) \rangle \, .
\end{eqnarray} 
Taking advantage of
the independence of fragmentation and initial kinematics once again,
we can separate the expectation values of the azimuths and apply
eq.~(\ref{nearsin}) and the approximation in the top line of
eq.~(\ref{neareq4}) for the fragmentation process:
\begin{eqnarray}
\langle p^2_{out,A} \rangle
&=& (\langle p^2_{T,assoc} \rangle 
- \langle p^2_{T,assoc} \sin^2 \Phi_{21} \rangle) \, \sigma_N
+ \langle p^2_{T,assoc} \sin^2 \Phi_{21} \rangle
\left( 1 - \sigma^2_N   \right)
\nonumber \\
&=& \langle p^2_{T,assoc} \rangle \, \sigma_N +
\langle p^2_{T,assoc} \sin^2 \Phi_{21} \rangle
\left( 1 - 2 \sigma^2_N \right)
\,\,\, .
\label{awayeq0}
\end{eqnarray}
Here we recognize that $\langle p^2_{T,assoc} \sin^2 \Phi_{21} \rangle$
is related to the intrinsic $k_T$ of the colliding
partons in the incoming protons, as discussed in Section 2. 
Recall that $p_{T2}$ and $\Phi_{21}$ are
not independent, and we wish to use eq.~(\ref{partonpc1})
to implement a partially correlated situation.
We express the corresponding term from eq.~(\ref{awayeq0}):
\begin{eqnarray}
{\langle p^2_{T,assoc} \sin^2 \Phi_{21} \rangle}& = &
              \frac{\langle p^2_{T,assoc} \rangle \, 
                     (\sigma^2_A - \sigma^2_N ) }
    {1 - 2 \sigma^2_N} \,\,\, . 
\label{complex1}
\end{eqnarray}
For simplicity we applied the top line of eq.~(\ref{neareq4}) 
to connect the 
away-side angular correlation  (characterized by the width $\sigma_A$)
with the quantity $\langle p^2_{out,A} \rangle$.
At larger values of $\sigma_A$ the approximation in the bottom line
of eq.~(\ref{neareq4})  leads to a more precise expression.
Further approximations are also available in the 
literature~\cite{jjPHENIX,jia04,jjvitev}.

In order to capture the correlation properties of the left-hand side
we need to return to parton level.
To separate the fragmentation effects for the associated hadron 
we take into account that
\begin{eqnarray}
p_{T,assoc} \cdot \cos \phi_{aj} &=& z_{assoc} \cdot p_{T2}  \,\,\,  .
\label{jetz}
\end{eqnarray}
Utilizing the angular features of fragmentation we take
expectation values separately, 
except for $p^2_{T2} \sin^2 \Phi_{21}$ as discussed above,
\be
\langle p^2_{T,assoc} \sin^2 \Phi_{21} \rangle
= \frac{\langle z_{assoc} \rangle^2} {\langle \cos^2 \phi_{aj} \rangle}
         \ \langle p^2_{T2} \sin^2 \Phi_{21} \rangle   \,\,\, .     
\label{eqfin2}
\ee
The last term on the right-hand side of eq.(\ref{eqfin2}) reminds us
of eq.~(\ref{partonpc1}) for a partially correlated 
situation with a dependence on a
parton-level $p_{T1}$. To express this with the transverse momentum of the 
trigger hadron (see Fig.~4) we use (compare eq.~(\ref{ztrig}))
\be
p_{T,trig} \cdot \cos \phi_{tj} &=& z_{trig} \cdot p_{T1} 
\label{jetza} \,\,\, . 
\ee
Combining eqs.(\ref{partonpc1}), (\ref{complex1}), 
(\ref{eqfin2}), and (\ref{jetza}),
we obtain
\begin{eqnarray}
\frac{\sigma^2_A -  \sigma^2_N}{1 - 2 \ \sigma^2_N}
&=& \frac{\langle z_{assoc} \rangle^2}
{\langle \cos^2 \phi_{aj} \rangle \ \langle p^2_{T,assoc} \rangle} 
\left[ \frac{\alpha}{4} p^2_{T1} + 2 \cdot \sigma^2_k \right] = \nonumber \\
&=& \frac{\langle z_{assoc} \rangle^2}
{\langle \cos^2 \phi_{aj} \rangle \ \langle p^2_{T,assoc} \rangle}
\left[ \frac{\alpha}{4} 
\frac{\langle \cos^2 \phi_{tj} \rangle}{\langle z_{trig} \rangle^2}
p^2_{T,trig} + 2 \cdot \sigma^2_k \right] \,\,\, .
\end{eqnarray}
If the azimuthal angles $\phi_{tj}$ and $\phi_{aj}$ are
independent and uncorrelated, then 
$\langle \cos^2 \phi_{tj} \rangle = \langle \cos^2 \phi_{aj} \rangle = 1/2$. 
In this case we get
\begin{eqnarray}
\frac{\sigma^2_A -  \sigma^2_N}{1 - 2 \ \sigma^2_N}
&=& \frac{2 \langle z_{assoc} \rangle^2}{\langle p^2_{T,assoc} \rangle}
\left[ \frac{ \alpha \, p^2_{T,trig}}{8 \langle z_{trig} \rangle^2}
 + 2 \sigma^2_k \right]  \,\,\, .
\end{eqnarray}
The message of this expression is that at a given $p_{T,trig}$,
measuring $\sigma_A$ and $\sigma_N$ in a $p_{T,assoc}$ window,  
and estimating $\langle z_{assoc} \rangle$, one can read off $\sigma_k$ from
the small $p_{T,trig}$ behavior of the combination on the
left-hand side, as the limit $p_{T,trig} \rightarrow 0$.
In that limit the wanted parton-level averaged transverse
momentum is
\begin{equation}
\langle \vert k_{Ty} \vert \rangle 
\equiv \sqrt{\frac{2 \sigma_k^2}{\pi}} 
= \frac{\sqrt{\langle p^2_{T,assoc} \rangle}}{\sqrt{\pi} \, 
\langle z_{assoc} \rangle}
\sqrt{\frac{\sigma^2_A -  \sigma^2_N}{1 - 2 \ \sigma^2_N}} \, \, \, .
\label{finsig}
\end{equation} 

We propose that the data on near and away-side correlations be
evaluated in this manner. The denominator $(1-2\sigma_N^2)$
will not approach zero according to available data~\cite{DNPRak},
but may be important numerically.
On the other hand, the data~\cite{DNPRak}
suggest, that  the application of a more precise approximation 
for the away side width (see 
the discussion around eq.~(\ref{neareq4}))
may be warranted~\cite{jjPHENIX,jia04,jjvitev}. 

\section{Summary}

We have shown in this paper that no information can be recovered on
the initial intrinsic transverse momentum distribution of partons in 
the proton if the transverse momenta of the outgoing partons are 
fully randomized.  However, proton-proton collisions display
partial randomization in the transverse plane, offering the
opportunity to extract the width of the intrinsic-$k_T$ distribution from
di-hadron correlations. We have found $p_{T,trig}$-dependent correlations
at ISR and RHIC energies, where the intrinsic-$k_T$ widths can be
extracted in the limit $p_{T,trig} \rightarrow 0$.
This procedure survives the complications introduced by fragmentation
and we propose a method to obtain the wanted initial width from hadron-hadron
correlations using the widths of the near and away-side peaks
of the hadron-hadron correlation functions.

At fix $p_{T,trig}$ the dependence of $\langle \vert k_{Ty} \vert \rangle$
on $p_{T,assoc}$ displayed in this paper, especially
eq.(\ref{finsig}), provides a new insight for recent experimental 
analyses (see e.g. Ref.~\cite{DNPRak,DNPJia}).
In particular, the denominator $(1-2\sigma_N^2)$ can be important
in eq.(\ref{finsig}). The interplay between $\langle p^2_{T, assoc} \rangle$
and $\langle z \rangle$ at high $p_{T}$ requires further study.

The present discussion was restricted to proton-proton 
collisions. The generalization for nuclear collisions will include
$k_T$-broadening from multiple scattering and possibly unequal 
widths when the colliding nuclei are of different species. 
As mentioned above, in $dAu$ (and $AuAu$) collisions 
more final-state interactions of the outgoing partons are expected.
Therefore, a difference in the variation of the $p_{T,trig}$ dependence 
of $\langle \vert k_{Ty} \vert \rangle$
between $pp$ and $dAu$ reactions could provide an alternative way to
separate initial and final-state interactions of partons. So far, 
the extracted widths from $dAu$
collisions are very similar to the $pp$ values at RHIC energy, 
but more detailed experiments and analyses may refine the situation.
In addition,
we expect that the $k_T$-widths used to describe single-particle spectra 
and the ones extracted from
hadron-hadron correlations can be brought into complete accord 
along these lines.


\begin{thebibliography}{99}
\bibitem{Fn}
We continue to use the notation
$k_{Ty}$ and $j_{Ty}$ for the component of the transverse vectors 
perpendicular to the direction of the jet axis and $p_{out,N(A)}$
for transverse vectors perpendicular to the trigger direction 
(see Fig. \ref{nearaway} and Refs~\cite{jjPHENIX,jia04}).
\end{thebibliography}
\end{document}